\documentclass{elsart}
\usepackage{amsmath}
\usepackage{graphicx}
\begin{document}
\begin{frontmatter}
\title{Multiply subtractive Kramers-Kr\"{o}nig relations for arbitrary-order harmonic generation susceptibilities}
\author{Valerio Lucarini}
\address{Via Panicale 10 50123 Firenze, Italy}
\author{Jarkko J. Saarinen and Kai-Erik Peiponen}
\address{Department of Physics, University of Joensuu, P.O. Box 111, FIN-80101 Joensuu, Finland}
\begin{abstract}
Kramers-Kr\"{o}nig (K-K) analysis of harmonic generation optical
data is usually greatly limited by the technical inability to
measure data over a wide spectral range. Data inversion for real
and imaginary part of $\chi^{n}(n\omega; \omega, ... ,\omega)$ can
be more efficiently performed if the knowledge of one of the two
parts of the susceptibility in a finite spectral range is
supplemented with a single measurement of the other part for a
given frequency. Then it is possible to perform data inversion
using only measured data and subtractive K-K relations. In this
paper multiply subtractive K-K relations are, for the first time,
presented for the nonlinear harmonic generation susceptibilities.
The applicability of the singly subtractive K-K relations are
shown using data for third-order harmonic generation
susceptibility of polysilane.
\end{abstract}
\end{frontmatter}

\section{Introduction}\label{sect1}

In linear optics Kramers-Kr\"{o}nig (K-K) relations and sum rules
for linear susceptibility, $\chi^{(1)}(\omega)$, constitute
well-established tools of fundamental importance in the analysis
of linear optical data, since they relate refractive and
absorptive phenomena and provide the possibility to check the
self-consistency of experimental or model-generated data
\cite{Landau,Kaibook,Altarelli1,Altarelli2}. The foundation of
these fundamental integral relations lies in the principle of
causality in the response of the matter to the external radiation,
the conceptual bridge being established by the Titchmarsh's
theorem~\cite{Nussenzveig}.

Much effort has been placed in a very wide and detailed
theoretical and experimental investigation of nonlinear optical
processes due to their huge scientific and technological
relevance. Apart from results relative to specific cases
\cite{Price,Kogan,Caspers}, a more modern approach to the problem
of framing in a coherent and general fashion dispersion theory of
the nonlinear susceptibilities has started recently dating back to
the late 1980s \cite{Kai1,Kai2} and early 1990s
\cite{Bassani1,Chernyak,Rapapa}.

In the case of harmonic wave generation, which is probably the
single most representative process of all the nonlinear optical
phenomena, only recently a complete formulation of general K-K
relations and sum rules for $n$th order harmonic generation
susceptibility $\chi^{(n)}(n\omega; \omega, \cdots ,\omega)$ has
been obtained \cite{ValerioHm,ValerioH,jarkko}. Unfortunately,
even now there are relatively few studies that report on
independent measurements of the real and imaginary parts of the
harmonic generation susceptibilities
\cite{Torruellas1,Torruellas2}, and on the validity of K-K
relations in nonlinear experimental data inversion \cite{Kishida}.

Characteristic integral structure of K-K relations requires the
knowledge of the spectrum at a semi-infinite angular frequency
range. Unfortunately, in practical spectroscopy only finite
spectral range can be measured. The technical problem to measure
nonlinear optical spectrum at relatively wide energy range is
probably the single most important reason why experimental
research in this field has been depressed for a long time.
Fortunately, recent development of dye lasers is very promising.
However, at the moment such lasers seem to have relevance in
nonlinear optical spectroscopy for relatively low order nonlinear
processes.

In the context of linear optics singly \cite{Ahrenkiel} (SSKK) and
multiply \cite{Palmer} subtractive Kramers-Kr\"{o}nig (MSKK)
relations have been proposed in order to relax the limitations due
to finite-range data. As far as we know, they have never been
proposed for nonlinear susceptibilities and especially for
harmonic generation susceptibilities. The idea behind the
subtractive Kramers-Kr\"{o}nig technique is that the inversion of
the real (imaginary) part of $\chi^{(n)}(n\omega; \omega, \cdots
,\omega)$ can be greatly improved if we have one or more anchor
points, i.e. a single or multiple measurement of the imaginary
(real) part for a set of frequencies. In such a case we give
general expressions for multiply subtractive K-K relations having
a faster convergence, thus decreasing the error due to the
inevitable finite spectral range.

This paper is organized as follows. In Section \ref{sect2} we give
the expressions for multiply subtractive K-K relations for
$\chi^{(n)}(n\omega; \omega, \cdots ,\omega)$ and in Section
\ref{sect3} we present application of SSKK on experimental data of
third-order harmonic generation susceptibility of polysilane.
Finally, in Section \ref{sect4} we set our conclusions.

\section{Multiply subtractive K-K relations for $\chi^{(n)}(n\omega; \omega, ... ,\omega)$}\label{sect2}

The analysis of the holomorphic \cite{Kaibook} properties of the
$n$th order harmonic generation susceptibility, which
intrinsically derive from the principle of causality in the
nonlinear response function of the matter \cite{Milonni}, allows
the derivation of the following Hilbert transform \cite{Bassani1}:
\begin{equation}\label{k1}
\imath \pi \chi^{(n)}(n\omega'; \omega', \cdots
,\omega')=\wp\int_{-\infty}^{\infty}\frac{\chi^{(n)}(n\omega;
\omega, \cdots ,\omega)}{\omega'-\omega}{\rm{d}}\omega,
\end{equation}
where $\wp$ indicates the Cauchy principal part integration. With
the aid of the symmetry relation
\begin{equation}\label{k2}
\chi^{(n)}(n\omega; \omega, \cdots ,\omega)=[\chi^{(n)}(-n\omega;
-\omega, \cdots ,-\omega)]^{\ast}
\end{equation}
with ($^{\ast}$) denoting the complex conjugation, we obtain the
following K-K relations for the real and imaginary parts:
\begin{equation}\label{k3}
\Re\{\chi^{(n)}(n\omega')\}=\frac{2}{\pi}\wp\int_{0}^{\infty}\frac{\omega\Im\{\chi^{(n)}(n\omega)\}}{\omega^{2}-\omega'^{2}}{\rm{d}}\omega,
\end{equation}
\begin{equation}\label{k4}
\Im\{\chi^{(n)}(n\omega')\}=-\frac{2\omega'}{\pi}\wp\int_{0}^{\infty}\frac{\Re\{\chi^{(n)}(n\omega)\}}{\omega^{2}-\omega'^{2}}{\rm{d}}d\omega,
\end{equation}
where, for the sake of clarity, we denote
$\chi^{(n)}(n\omega;\omega, \cdots ,\omega)$ simply by
$\chi^{(n)}(n\omega)$. The independent dispersion relation
(\ref{k3}) in principle allows us to compute the real part of the
susceptibility once we know the imaginary part for all frequencies
and {\textit{vice versa}}.

Palmer $et$ $al$. \cite{Palmer} have studied multiply subtractive
K-K analysis in the case of phase retrieval problems related to
linear reflection spectroscopy. Here their results are generalized
to hold for holomorphic nonlinear susceptibilities. Unfortunately,
Palmer $et$ $al$. \cite{Palmer} presented MSKK only for the phase
angle (imaginary part of the linear reflectance). Here we extend
their theory to hold both for the real and imaginary parts of the
arbitrary-order harmonic generation susceptibility. With the aid
of mathematical induction (see appendix A in ref. \cite{Palmer})
we can derive the multiply subtractive K-K relation for the real
and imaginary parts as follows:
\begin{equation}\label{k5}
\begin{split}
\Re\{\chi&^{(n)}(n\omega')\}\\&=\left[\frac{(\omega'^2-\omega_2^2)(\omega'^2-\omega_3^2)\cdots(\omega'^2-\omega_Q^2)}{(\omega_1^2-\omega_2^2)(\omega_1^2-\omega_3^2)\cdots(\omega_1-\omega_Q^2)}\right]\Re\{\chi^{(n)}(n\omega_1)\}+\cdots\\
&+\left[\frac{(\omega'^2-\omega_1^2)\cdots(\omega'^2-\omega_{j-1}^2)(\omega'^2-\omega_{j+1}^2)\cdots(\omega'^2-\omega_Q^2)}{(\omega_j^2-\omega_1^2)\cdots(\omega_j^2-\omega_{j-1}^2)(\omega_j^2-\omega_{j+1}^2)\cdots(\omega_j-\omega_Q^2)}\right]\Re\{\chi^{(n)}(n\omega_j)\}+\cdots\\
&+\left[\frac{(\omega'^2-\omega_1^2)(\omega'^2-\omega_2^2)\cdots(\omega'^2-\omega_{Q-1}^2)}{(\omega_Q^2-\omega_1^2)(\omega_Q^2-\omega_2^2)\cdots(\omega_Q-\omega_{Q-1}^2)}\right]\Re\{\chi^{(n)}(n\omega_Q)\}\\
&+\frac{2}{\pi}\left[(\omega'^2-\omega_1^2)(\omega'^2-\omega_2^2)\cdots(\omega'^2-\omega_Q^2)\right]\wp\int_0^{\infty}\frac{\omega\Im\{\chi^{(n)}(n\omega)\}\d\omega}{(\omega^2-\omega'^2)\cdots(\omega^2-\omega_Q^2)},
\end{split}
\end{equation}
\begin{equation}\label{k6}
\begin{split}
\frac{\Im\{\chi^{(n)}(n\omega')\}}{\omega'}&=\left[\frac{(\omega'^2-\omega_2^2)(\omega'^2-\omega_3^2)\cdots(\omega'^2-\omega_Q^2)}{(\omega_1^2-\omega_2^2)(\omega_1^2-\omega_3^2)\cdots(\omega_1-\omega_Q^2)}\right]\frac{\Im\{\chi^{(n)}(n\omega_1)\}}{\omega_1}+\cdots\\
&+\left[\frac{(\omega'^2-\omega_1^2)\cdots(\omega'^2-\omega_{j-1}^2)(\omega'^2-\omega_{j+1}^2)\cdots(\omega'^2-\omega_Q^2)}{(\omega_j^2-\omega_1^2)\cdots(\omega_j^2-\omega_{j-1}^2)(\omega_j^2-\omega_{j+1}^2)\cdots(\omega_j-\omega_Q^2)}\right]\frac{\Im\{\chi^{(n)}(n\omega_j)\}}{\omega_j}+\cdots\\
&+\left[\frac{(\omega'^2-\omega_1^2)(\omega'^2-\omega_2^2)\cdots(\omega'^2-\omega_{Q-1}^2)}{(\omega_Q^2-\omega_1^2)(\omega_Q^2-\omega_2^2)\cdots(\omega_Q-\omega_{Q-1}^2)}\right]\frac{\Im\{\chi^{(n)}(n\omega_Q)\}}{\omega_Q}\\
&-\frac{2}{\pi}\left[(\omega'^2-\omega_1^2)(\omega'^2-\omega_2^2)\cdots(\omega'^2-\omega_Q^2)\right]\wp\int_0^{\infty}\frac{\Re\{\chi^{(n)}(n\omega)\}\d\omega}{(\omega^2-\omega'^2)\cdots(\omega^2-\omega_Q^2)}.
\end{split}
\end{equation}
Here $\omega_j$ with $j=1, \cdots ,Q$ denote the anchor points.
Note that the anchor points in Eqs. (\ref{k5}) and (\ref{k6}) need
not to be the same. We observe that the integrands of Eqs.
(\ref{k5}) and (\ref{k6}) have remarkably faster asymptotic
decrease, as a function of angular frequency, than the
conventional K-K relations given by Eqs. (\ref{k3}) and
(\ref{k4}). This can be observed by comparing the integrands of
K-K and MSKK relations since the convergence of $Q$-times
subtracted K-K relations is proportional to $\omega^{-(2n+2+2Q)}$
whereas the conventional K-K relations decrease proportional to
$\omega^{-(2n+2)}$. Therefore, it is expected that the limitations
related to the presence of an experimentally unavoidable finite
frequency range are thus relaxed, and the precision of the
integral inversions is then enhanced.

Before proceeding we wish to remark that MSKK relations can also
be written for all the moments
$\omega'^{2\alpha}[\chi^{(n)}(n\omega'; \omega',...,\omega')]^k$
with $0\leq\alpha\leq k(n+1)$, where $\alpha$ and $k$ are
integers. Such functions play an important role in the context of
sum rules for arbitrary-order harmonic generation susceptibilities
\cite{ValerioHm,ValerioH,jarkko}.

Palmer $et$ $al$. \cite{Palmer} discussed how the anchor points
should be chosen inside the measured spectral range. It is well
known that Chebyshev polynomials have great importance in
minimizing errors in numerical computations \cite{arfken}.
According to Palmer $et$ $al$. \cite{Palmer} accurate data
inversion is possible when the anchor points are chosen near to
the zeros of the $Q$th order Chebyshev polynomial of the first
kind. In linear optical spectroscopy it is usually easy to get
information of the optical constants at various anchor points.
However, in the field of nonlinear optics it is difficult to
obtain the real and imaginary parts of the nonlinear
susceptibility at various anchor points. Therefore, in the present
study we wish to emphasize that even a single anchor point reduces
the errors caused by finite spectral range in data inversion of
nonlinear optical data. Then the choice of the location of the
anchor point is not critical as concerns the coincidence of the
zero of the Chebyshev polynomial. Furthermore, the Chebyshev zeros
accumulate at the ends of the data interval. This is the reason
why the anchor point is chosen near to one end of the data
interval. For one anchor point, say at frequency $\omega_1$, we
obtain from Eqs. (\ref{k5}) and (\ref{k6}) the following singly
subtractive K-K relations
\begin{eqnarray} \label{k7}
\lefteqn{\Re\{\chi^{(n)}(n\omega')\}-
\Re\{\chi^{(n)}(n\omega_{1})\}{} } \nonumber\\ & &
{}=\frac{2(\omega'^{2}-\omega_{1}^{2})}{\pi}\wp\int_{0}^{\infty}\frac{\omega\Im\{\chi^{(n)}(n\omega)\}}{(\omega^{2}-\omega'^{2})(\omega^{2}-\omega_{1}^{2})}{\rm{d}}\omega,
\end{eqnarray}
\begin{eqnarray} \label{k8}
\lefteqn{\omega'^{-1}\Im\{\chi^{(n)}(n\omega')\}-
\omega_{1}^{-1}\Im\{\chi^{(n)}(n\omega_{1})\}{} } \nonumber\\ & &
{}=-\frac{2(\omega'^{2}-\omega_{1}^{2})}{\pi}\wp\int_{0}^{\infty}\frac{\Re\{\chi^{(n)}(n\omega)\}}{(\omega^{2}-\omega'^{2})(\omega^{2}-\omega_{1}^{2})}{\rm{d}}\omega,
\end{eqnarray}
which are used for the experimental data analysis

\section{Application of singly subtractive K-K relations to experimental data of $\chi^{(3)}(3\omega; \omega,\omega
,\omega)$ on polysilane} \label{sect3}

Here we apply singly subtractive K-K relations for real data, in
order to prove their effective relevance. We consider the
experimental values of the real and imaginary part of the
nonlinear susceptibility of third-order harmonic wave generation
on polysilane, obtained by Kishida $et$ $al$. \cite{Kishida}; for
both $\Re\{\chi^{(3)}(3\omega; \omega,\omega ,\omega)\}$ and
$\Im\{\chi^{(3)}(3\omega; \omega,\omega ,\omega)\}$, which come
from independent measurements, the energy range is $0.4-2.5$ eV.

First we consider only data ranging from $0.9$ to $1.4$ eV, in
order to simulate a low-data availability scenario, and compare
the quality of the data inversion obtained with the conventional
K-K and SSKK relations within this energy range. This interval
constitutes a good test since it contains the most relevant
feature of both parts of the susceptibility. However, a lot of the
spectral structure is left outside the interval and the asymptotic
behavior is not established for either parts. Therefore, no plain
optimal conditions for optical data inversion are established.

In Fig. $1$ we show the results obtained for the real part of the
third-order harmonic generation susceptibility. The solid line in
Fig. 1 represents the experimental data. The dashed curve in Fig.
1, which was calculated by using conventional K-K relation by
truncating integration of (\ref{k3}) consistently gives a poor
match with the actual line. On the contrary, we obtain a better
agreement with a single anchor point located at $\omega _{1}=0.9$
eV, which is represented by dotted line in Fig. 1. SSKK and
measured data for the real part of the susceptibility are almost
undistinguishable up to $1.3$ eV.

In Fig. $2$ similar calculations as above are shown but for the
imaginary part of the nonlinear susceptibility. In this case the
anchor point is located at $\omega_{1}=1$ eV. From Fig. 2 we
observe that the precision of the data inversion is dramatically
better by using SSKK instead of the conventional K-K relations.
The presence of the anchor point greatly reduces the errors of the
estimation performed with the conventional K-K relations in the
energy range $0.9-1.4$ eV.

\section{Conclusions}\label{sect4}

The extrapolations in K-K analysis, such as the estimation of the
data beyond the measured spectral range, can be a serious source
of errors \cite{Kai3,Aspnes}. Recently, King \cite{King2}
presented an efficient numerical approach to the evaluation of K-K
relations. Nevertheless, the problem of data fitting is always
present in regions outside the measured range.

In this paper we proposed how an independent measurement of the
unknown part of the complex third-order nonlinear susceptibility
for a given frequency relaxes the limitations imposed by the
finiteness of the measured spectral range, since in the obtained
SSKK relations faster asymptotic decreasing integrands are
present. SSKK relations can provide a reliable data inversion
procedure based on using {\textit{measured data only}}. We
demonstrated that SSKK relations yield more precise data
inversion, using only a single anchor point, than conventional K-K
relations.

Naturally it is possible to exploit also MSKK if higher precision
is required. However, the measurement of multiple anchor may be
experimentally tedious. Finally, we remark that MSKK relations are
valid for all holomorphic nonlinear susceptibilities of
arbitrary-order. As an example of such holomorphic third-order
nonlinear susceptibilities we mention those related to pump and
probe nonlinear processes (see details of the various expressions
in Ref. \cite{Valerio}). Unfortunately, the degenerate
arbitrary-order nonlinear susceptibility is a meromorphic function
\cite{Peiponen02} and MSKK cannot be applied.

\section*{Acknowledgments}
The authors would like to express their cordial thanks to Dr.
Hideo Kishida (Department of Advanced Materials Science,
University of Tokyo) and Dr. Takao Koda (Professor emeritus,
University of Tokyo) for providing the measured optical data on
polysilane. One of the authors (JJS) wishes to thank the Nokia
foundation for financial support.

\newpage

\section*{Figure captions}

Figure 1: Efficacy of SSKK vs. K-K relations in retrieving
$\Re\{\chi^{(3)}(3\omega; \omega, \omega, \omega)\}$.
\par
Figure 2: Efficacy of SSKK vs. K-K relations in retrieving
$\Im\{\chi^{(3)}(3\omega; \omega, \omega, \omega)\}$.

\newpage
\pagestyle{empty}

\begin{figure}
\centering
\includegraphics[angle=270,width= 16 cm]{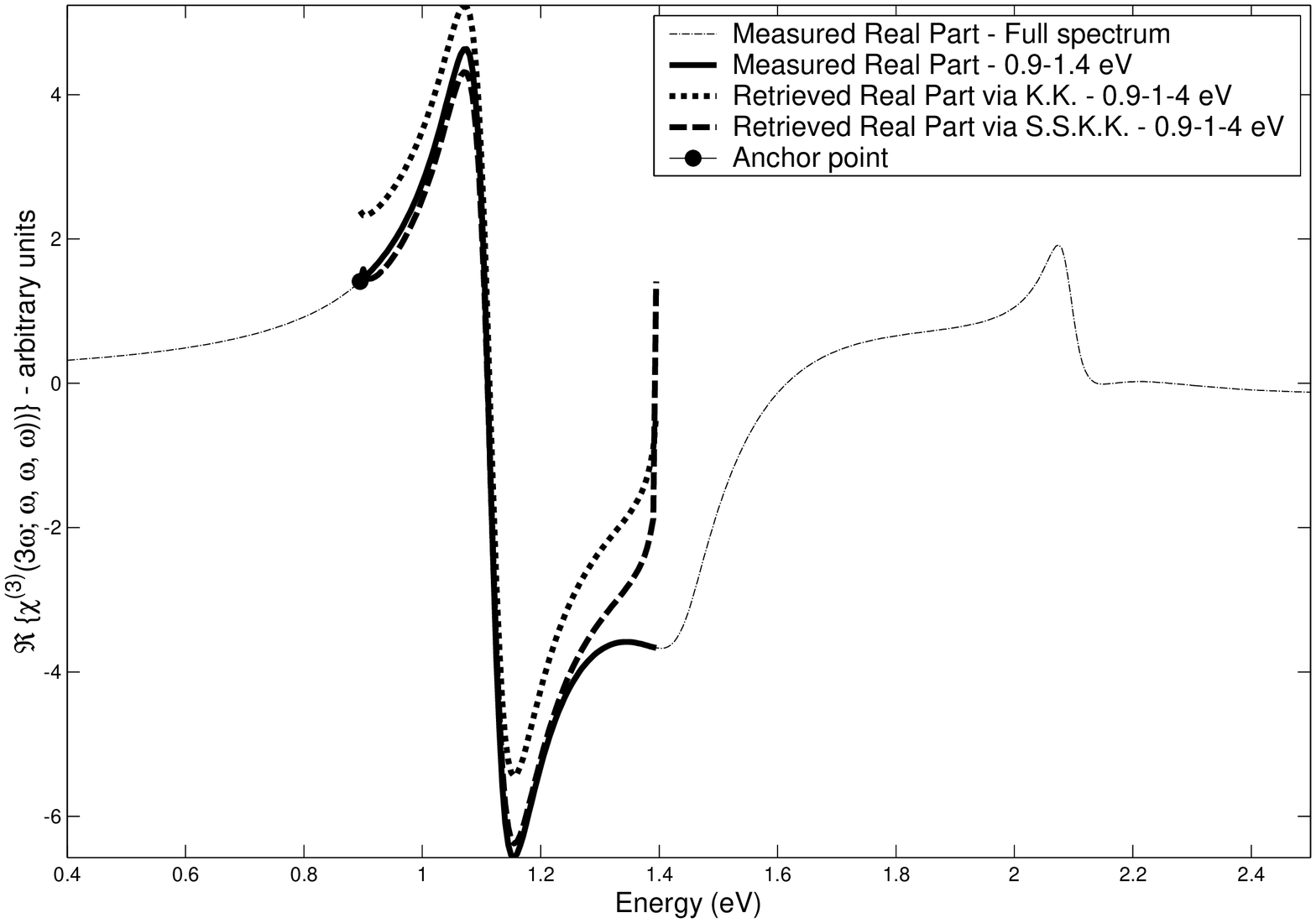}
\end{figure}

\vfill

\begin{figure}
\begin{center}
{\bf{Figure 1:}} Lucarini, Saarinen, and Peiponen.
\end{center}
\end{figure}

\newpage
\pagestyle{empty}

\begin{figure}
\centering
\includegraphics[angle=270,width= 16 cm]{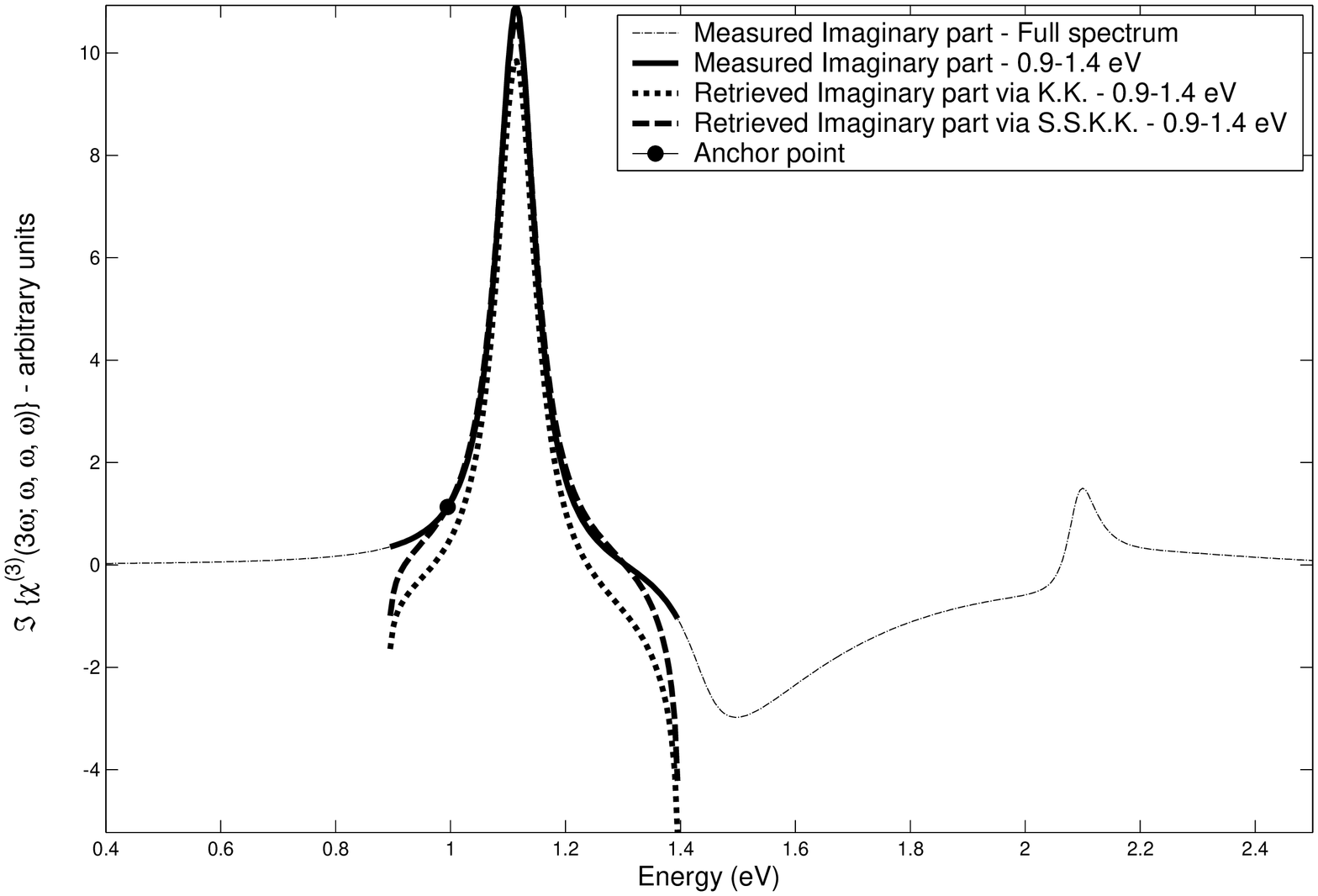}
\end{figure}

\vfill

\begin{figure}
\begin{center}
{\bf{Figure 2:}} Lucarini, Saarinen, and Peiponen.
\end{center}
\end{figure}

\end{document}